\documentclass{PoS}

\usepackage[T1]{fontenc}
\usepackage[latin9]{inputenc}

\usepackage{amsmath,amssymb,amsthm,amsfonts}
\usepackage{graphicx}
\usepackage{xcolor}

\usepackage{tabularx,booktabs}

\newcommand{\eprintlink}[2]{\href{#1}{\color{blue}#2}}

\definecolor{links}{rgb}{0,0.3,0}
\newcommand{\Graph}[2][1.0]{
\vcenter{\hbox{\includegraphics[scale=#1]{#2}}}}

\title{The number of master integrals as Euler characteristic}

\ShortTitle{The number of master integrals as Euler characteristic}

\author{Thomas Bitoun\\
        Mathematical Institute, University of Oxford, Oxford OX2 6GG, UK\\
        E-mail: \email{tbitoun@gmail.com}}
\author{Christian Bogner\\ Institut f\"ur Physik, Humboldt-Universit\"at zu Berlin,
D - 10099 Berlin, Germany\\
 E-mail: \email{bogner@math.hu-berlin.de}}
\author{Ren\'e Pascal Klausen\\ Institut f\"ur Physik, Humboldt-Universit\"at zu Berlin,
D - 10099 Berlin, Germany\\
        E-mail: \email{klausen@physik.hu-berlin.de}}
\author{\speaker{Erik Panzer}\\
        All Souls College, Oxford OX1 4AL, UK\\
        E-mail: \email{erik.panzer@all-souls.ox.ac.uk}}

\abstract{We give a brief introduction to a parametric approach for the derivation of shift relations between Feynman integrals and a result on the number of master integrals. The shift relations are obtained from parametric annihilators of the Lee-Pomeransky polynomial $\mathcal{G}$. By identification of Feynman integrals as multi-dimensional Mellin transforms, we show that this approach generates every shift relation. Feynman integrals of a given family form a vector space, whose finite dimension is naturally interpreted as the number of master integrals. This number is an Euler characteristic of the polynomial $\mathcal{G}$.}

\FullConference{Loops and Legs in Quantum Field Theory (LL2018)\\
		29 April 2018 - 04 May 2018\\
		St. Goar, Germany}

\begin{document}

\section{Introduction}

We consider families of Feynman integrals in $d$ dimensions
\begin{equation}
\mathcal{I}(\nu_{1},...,\nu_{N})=\left(\prod_{j=1}^{L}\frac{d^{d}l_{j}}{i\pi^{\frac{d}{2}}}\right)\prod_{a=1}^{N}D_{a}^{-\nu_{a}}\label{eq:family}
\end{equation}
where the $D_{a}$ are at most quadratic forms in the loop momenta
$l_{1},...,l_{L}$ and in external momenta. Such $D_{a}$ include
the usual inverse Feynman propagators and irreducible scalar products.
Relations between members of such an integral family which differ
from each other by integer shifts of the indices $\nu_{1},...,\nu_{N}$
are known as shift relations. They are usually obtained from the classical
integration-by-parts (IBP) method \cite{CheTka,Tkachov:calculability-4loop} by insertion of
differentiations with respect to the loop momenta under the integral
sign. Alternative approaches to the derivation of shift relations
were proposed in the past \cite{Baikov:ExplicitSolutions-nloop, Lee:ModernTechniques, Tkachov:WhatsNext}, based on parametric representations
of Feynman integrals and the insertion of corresponding differential
operators. 

Here we elaborate on one of these alternative approaches which was suggested by
Lee \cite{Lee:ModernTechniques}. Starting point of the method is the Lee-Pomeransky polynomial $\mathcal{G}$ of the Feynman graph which is simply the sum of the two Symanzik polynomials. A certain representation of Feynman integrals in terms of Schwinger parameters \cite{LeePom} is recognized to be a multi-dimensional
Mellin transform of $\mathcal{G}^{-\frac{d}{2}}$. The correspondence between an inserted differential
operator and the resulting shift relation is directly obtained from
properties of the Mellin transform. From the invertibility of Mellin
transforms it is furthermore clear, that the approach generates every
existing shift relation.

The inserted differential operators, which act on $\mathcal{G}^{s}$ where $s=-\frac{d}{2}$ are known in the mathematical literature as the $s$-parametric
annihilators of this polynomial. They form an ideal which can be obtained
algorithmically by use of computer algebra systems such as \textsc{Singular}
\cite{Andetal, Singular}.

Making use of the fact that this ideal defines a holonomic D-module,
we arrive at a statement on the number of master integrals, which
is the main result of our work. Applying a theorem of Loeser and Sabbah
\cite{LoeserSabbah:IrredTore, LoeserSabbah:IrredToreII}, we identify the number of master integrals of a family
of Feynman integrals with an Euler characteristic, which can be computed
explicitly from the corresponding graph polynomial by various methods.

This presentation serves as a quick introduction to these aspects,
while all proofs, many examples and further details are provided in
\cite{Bitetal}. Our text is structured as follows: In section \ref{sec:Mellin} we write the Feynman integrals as Mellin transforms and recall some relevant properties of the latter. In section \ref{sec:Ann} we introduce algebras of the inserted differential operators and shift operators and demonstrate the correspondence between them. Section \ref{sec:Euler} reviews our main result which expresses the number of master integrals as an Euler characteristic. In the final section we state our conclusions and point out some open questions.

\section{Feynman integrals as Mellin transforms}\label{sec:Mellin}

Throughout this presentation we use the notation
\[
s:=-\frac{d}{2}\in\mathbb{C}
\]
where $d$ is the space-time dimension. For each $D_{a}$ in the integrand of eq. \ref{eq:family} we introduce
a Schwinger parameter $x_{a}$ and by the usual decomposition 
\[
\sum_{a=1}^{N}x_{a}D_{a}=-\sum_{i,j=1}^{L}\Lambda_{ij}l_{i}l_{j}+\sum_{i=1}^{L}2Q_{i}l_{i}+J
\]

into terms quadratic, linear and constant in the loop momenta, we
obtain the well-known Symanzik polynomials 
\[
\mathcal{U}:=\det\Lambda \;\;\;\; \textrm{ and }\;\;\;\;\mathcal{F}:=\mathcal{U}\left(Q^{T}\Lambda^{-1}Q+J\right)
\]
where the $L\times L$-matrix $\Lambda,$ the vector $Q$ and the
scalar $J$ are defined by the above decomposition. One way of using
these polynomials to express $\mathcal{I}(\nu)$ as an integral over
the Schwinger parameters is 
\[
\mathcal{I}(\nu)=\frac{\Gamma\left(-s\right)}{\Gamma\left(-s-\omega\right)}\left(\prod_{i=1}^{N}\int_{0}^{\infty}\frac{x_{i}^{\nu_{i}-1}dx_{i}}{\Gamma(\nu_{i})}\right)\mathcal{G}^s
\]
as advocated in \cite{LeePom}, where $\mathcal{G}=\mathcal{U}+\mathcal{F}$
and $\omega=sL+\sum_{i=1}^{N}\nu_{i}.$ It is useful to
view this representation as a Mellin transform. For $\nu=\left(\nu_{1},...,\nu_{N}\right)\in\mathbb{C}^{N}$
we define the twisted (multi-dimensional) Mellin transform of a function
$f:\,\mathbb{R}_{+}^{N}\rightarrow\mathbb{C}$ as 
\begin{equation}
\mathcal{M}\{f\}(\nu):=\left(\prod_{i=1}^{N}\int_{0}^{\infty}\frac{x_{i}^{\nu_{i}-1}dx_{i}}{\Gamma(\nu_{i})}\right)f\left(x_{1},...,x_{N}\right)\label{eq:Mellin transform}
\end{equation}
assuming conditions such that the integral exists. The Feynman integral becomes
\begin{equation}
\mathcal{I}(\nu)=\frac{\Gamma\left(-s\right)}{\Gamma\left(-s-\omega\right)}\mathcal{M}\left\{ \mathcal{G}^s \right\}(\nu)\label{eq:Feynman integral as Mellin transformation}
\end{equation}
and it will be convenient to derive shift relations for 
\[
\tilde{\mathcal{I}}(\nu):=\mathcal{M}\left\{ \mathcal{G}^s\right\} (\nu)
\]
at first and to restore the gamma factors afterwards.

It is crucial for our discussion to consider the Feynman integrals
as functions of complex variables $s,\nu_{1},...,\nu_{N}$ instead
of restricting the indices only to integer values. As a consequence,
the precise conditions under which the
integrals of eq. \ref{eq:Mellin transform} are defined are not relevant for our discussion. As every
Feynman integral extends to a unique meromorphic function of complex
$s,\nu_{1},...,\nu_{N}$ by analytic continuation \cite{Speer:GeneralizedAmplitudes, Speer:SingularityStructureGenericFeynmanAmplitudes}, every
shift relation which can be established in some domain where the integrals
exist will also be satisfied by the unique extensions. Therefore it
is sufficient to know that such a domain always exists.

The way in which indices are shifted by the insertion of differential
operators follows from the properties 
\begin{align}
\mathcal{M}\left\{ \alpha f+\beta g\right\} (\nu) & =\alpha\mathcal{M}\{f\}(\nu)+\beta\mathcal{M}\{g\}(\nu),\label{eq:Mt1}\\
\mathcal{M}\left\{ x_{i}f\right\} (\nu) & =\nu_{i}\mathcal{M}\{f\}\left(\nu+e_{i}\right)=\left(\hat{{\bf i}}^{+}\mathcal{M}\{f\}\right)(\nu),\label{eq:Mt2}\\
\mathcal{M}\left\{ \partial_{i}f\right\} (\nu) & =-\mathcal{M}\{f\}\left(\nu-e_{i}\right)=-\left({\bf i}^{-}\mathcal{M}\{f\}\right)(\nu)\label{eq:Mt3}
\end{align}
of the twisted Mellin transform, with $\alpha,\beta\in\mathbb{C}$
and where $e_{i}$ denotes the $i$-th unit vector. Here $\hat{{\bf i}}^{+}$
and ${\bf i}^{-}$ are the usual shift operators defined by 
\begin{align*}
\left({\bf i}^{-}F\right)(\nu) & :=F\left(\nu-e_{i}\right),\\
\left(\hat{{\bf i}}^{+}F\right)(\nu) & :=\nu_{i}F\left(\nu+e_{i}\right)
\end{align*}
and we furthermore use ${\bf n}_{i}:=\hat{{\bf i}}^{+}{\bf i}^{-}$
with $\left({\bf n}_{i}F\right)(\nu)=\nu_{i}F(\nu).$

\section{Annihilators and shift relations}\label{sec:Ann}

We consider two algebras of operators: Differential operators in the
Weyl algebra 
\[
A^{N}[s]:=\mathbb{C}[s]\left\langle x_{1},...,x_{N},\partial_{1},...,\partial_{N}\left|\left[\partial_{i},x_{j}\right]=\delta_{ij}\right.\right\rangle ,
\]
acting on integrands, and operators in the shift algebra 
\[
S^{N}[s]:=\mathbb{C}[s]\left\langle \hat{{\bf 1}}^{+},...,\hat{{\bf N}}^{+},{\bf 1}^{-},...,{\bf N}^{-}\left|\left[-{\bf j}^{-},\hat{{\bf i}}^{+}\right]=\delta_{ij}\right.\right\rangle ,
\]
acting on integrals. Similar to the IBP method, we insert differential
operators in the integrand, such that the integral vanishes. In contrast
to the IBP method, the inserted differential operators are precisely
the ones which make the integrand vanish. 

For a given integral $\tilde{\mathcal{I}}(\nu)=\mathcal{M}\left\{ \mathcal{G}^s\right\} (\nu)$
we consider the annihilators $P\in A^{N}[s]$ of $\mathcal{G},$ defined
by 
\[
P\mathcal{G}^s=0.
\]
According to eqs. \ref{eq:Mt1}-\ref{eq:Mt3} we obtain for each $P$
a shift operator, say $\mathcal{M}\{P\},$ by substituting 
\[
x_{i}\longmapsto\hat{{\bf i}}^{+},\;\partial_{i}\longmapsto-{\bf i}^{-},\;x_{i}\partial_{i}\longmapsto-{\bf n}_{i}.
\]
For each annihilator $P$ we have 
\[
\mathcal{M}\left\{ P\mathcal{G}^s\right\} (\nu)=\mathcal{M}\{P\}\mathcal{M}\left\{ \mathcal{G}^s\right\} (\nu)=0,
\]
where the last equality is the desired shift relation. 

\textbf{Example:} For the graph polynomial 
\[
\mathcal{G}=x_{1}+x_{2}-p^{2}x_{1}x_{2}
\]
of the one-loop graph with two propagators and external momentum $p,$
we find an annihilator 
\[P=-p^{2}\left(s-x_{1}\partial_{1}+1\right)x_{1}+\left(s-x_{1}\partial_{1}-x_{2}\partial_{2}\right)
\]
whose
shift operator 
\[
\text{\ensuremath{\mathcal{M}}}\{P\}=-p^{2}\left(s+{\bf n}_{1}+1\right)\hat{{\bf 1}}^{+}+\left(s+{\bf n}_{1}+{\bf n}_{2}\right)
\]
leads to the relation 
\[-p^{2}\nu_{1}\tilde{\mathcal{I}}\left(\nu_{1}+1,\nu_{2}\right)=-\frac{s+\nu_{1}+\nu_{2}}{s+\nu_{1}+1}\tilde{\mathcal{I}}\left(\nu_{1},\nu_{2}\right).
\]
After re-introducing the gamma-factors of eq. \ref{eq:Feynman integral as Mellin transformation}
we obtain the shift relation 
\[
-p^{2}\nu_{1}\mathcal{I}\left(\nu_{1}+1,\nu_{2}\right)=\frac{\left(s+\nu_{1}+\nu_{2}\right)\left(2s+\nu_{1}+\nu_{2}+1\right)}{s+\nu_{1}+1}\mathcal{I}\left(\nu_{1},\nu_{2}\right).
\]

The Mellin transform is invertible and defines a bijection between
the annihilators and the shift relations. In other words, every shift
relation between Feynman integrals of a given family can be obtained
from an annihilator of the corresponding graph polynomial $\mathcal{G}.$
The annihilators of a given $\mathcal{G}$ form the ideal $\textrm{Ann}_{A^{N}[s]}\left(\mathcal{G}^s\right)$
whose generators can be derived automatically (see \cite{Andetal, Oak, OakTak}). In this
sense, all annihilators are in principle available. 

\section{The number of master integrals}\label{sec:Euler}

As an application of the above approach and main result of our
work, we present a statement on the number of master integrals. The
term ``number of master integrals'' is used for various quantities
in the literature. We propose the following definition: Let $V_{\mathcal{G}}$
be the vector space of all Feynman integrals associated to $\mathcal{G}$
over the field $\mathbb{C}(s,\nu)$, that is 
\[
V_{\mathcal{G}}:=\sum_{n\in\mathbb{Z}^{N}}\mathbb{C}(s,\nu)\cdot\mathcal{M}\left\{ \mathcal{G}^s\right\} (\nu+n).
\]
Then the number of master integrals is the dimension of this vectorspace
\[
\mathfrak{C}\left(\mathcal{G}\right):=\dim_{\mathbb{C}(s,\nu)}V_{\mathcal{G}}.
\]
In the literature, Feynman integrals are often restricted to integer
indices and many authors discard integrals where one or several indices
are zero, usually referred to as integrals of subtopologies, in their counting of master integrals. Our number $\mathfrak{C}\left(\mathcal{G}\right)$
instead includes the master integrals of subtopologies. It is also very
common to reduce the number of master integrals by use
of possible permutation symmetries of indices, which are not included
in our approach so far. Furthermore some authors discard master integrals which
can be expressed in terms of gamma functions or products of integrals.
For these reasons, our number will usually set an upper bound on other
countings.

The number $\mathfrak{C}\left(\mathcal{G}\right)$ is unambiguously defined and
can be computed exactly. As a first step, one notices that by use
of the inverse Mellin transform and of $\theta_{i}:=x_{i}\partial_{i}=\mathcal{M}^{-1}\{-\nu_{i}\},$
this number is the dimension of the space of the corresponding integrands,
i.e. 
\[
\mathfrak{C}\left(\mathcal{G}\right)=\dim_{\mathbb{C}(s,\theta)}\left(\mathbb{C}(s,\theta)\otimes_{\mathbb{C}[s,\theta]}A^{N}[s]\cdot\mathcal{G}^{s}\right).
\]
Defining $k:=\mathbb{C}(s),$ $R=k[\theta]$ and $F=\mathbb{C}(s,\theta)$, we can equivalently write
\[
\mathfrak{C}\left(\mathcal{G}\right)=\dim_F\left(F\otimes_R\mathcal{M}\right)
\]
where $\mathcal{M}=A_{k}^{N}\cdot\mathcal{G}^s$ is a module over the Weyl algebra $A_{k}^{N}:=A^N[s]\otimes_{\mathbb{C}[s]}k$ over $k$.
It is known from classical work of Bernstein \cite{Bernshtein:AnalyticContinuation} that the module $\mathcal{M}$ is holonomic.
It is for this reason that a theorem of Loeser and Sabbah \cite{LoeserSabbah:IrredTore, LoeserSabbah:IrredToreII} applies to our situation. This
theorem identifies the dimension of certain vector spaces defined
by $A_{k}^{N}$-modules with an Euler characteristic. We prove that
up to a sign this is in our case the topological Euler characteristic
of the complement of the graph hypersurface $x_{1}\cdot\cdot\cdot x_{N}\mathcal{G}=0$
in $\mathbb{C}^{N}.$ Our final result reads 
\[
(-1)^N \mathfrak{C}\left(\mathcal{G}\right) =  \chi \left(\mathbb{C}^N\backslash \left\{x_1
\cdot \cdot \cdot x_N\cdot \mathcal{G}=0\right\}\right) =  \chi \left(\left(\mathbb{C^\star}\right)^N\backslash \left\{\mathcal{G}=0\right\}\right).
\]
This result implies the finiteness of the number of master integrals which was proven for the first time in \cite{SmiPet}. More importantly, it expresses this number as a fundamental invariant. The Euler characteristic $\chi$ is the alternating sum of the dimensions of singular homology groups 
\[
\chi (X) = \sum_i (-1)^i \textrm{dim}H^i(X).
\]
This number can be computed by many tools. Firstly, it is very useful to consider the Euler characteristic as an invariant of
the equivalence class $\left[\mathcal{G}\right]$ of the variety $\mathbb{V}(\mathcal{G})=\left\{\mathcal{G}=0\right\}$ in the Grothendieck ring.
The latter is generated by isomorphism classes $[X]$ of varieties over $\mathbb{C}$, modulo relations $[X]=[X\backslash Z]+[Z]$ where $Z$ is a closed subvariety $Z\subset X$. As the Euler characteristic is compatible with this relation, one has 
\[
\chi(X)=\chi(X\backslash Z)+\chi(Z).
\]
Furthermore, for $A,B\in \mathbb{C}[x_1,...,x_{N-1}]$ and $A+x_N B$ a polynomial linear in $x_N$, one obtains   
\[
\chi \left(\mathbb{C^\star}\backslash \mathbb{V}(A+x_N B)\right) = -\chi \left(\mathbb{C^\star}\backslash \mathbb{V}(A\cdot B)\right).
\]
Applied to the varieties of Symanzik polynomials, these relations are very useful to reduce the computation of the Euler characteristic. In particular, for linearly reducible Feynman graphs (see \cite{Bro}), the entire computation can be done following an algorithm similar to \cite{Ste}. Many of our examples presented in \cite{Bitetal} were computed in this way. 

For the general purpose of computing the Euler characteristic, independent of properties like linear reducibility, the computer algebra system \textsc{Macaulay2} \cite{Macaulay2} provides the command \texttt{Euler} in the package \texttt{CharacteristicClasses}. We have used this command to compute the numbers in table \ref{tab:examples}. All of our results have been checked with the program \textsc{Azurite} \cite{GeorgoudisLarsenZhang:Azurite}.
	
\begin{table}
\begin{centering}	
\begin{tabular}{ccccc}
	Graph $G$ & $\Graph[0.8]{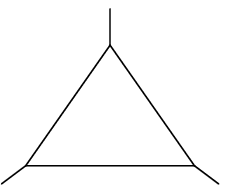}$ & $\Graph[0.55]{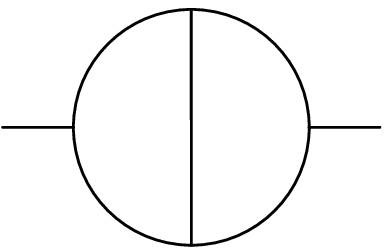}$ & $\Graph[0.8]{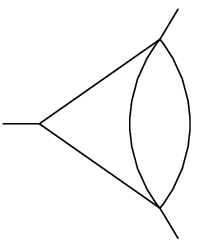}$ & $\Graph[0.8]{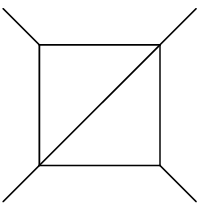}$ \\

$\mathfrak{C}{G}$ massless & 4 & 3 & 4 & 20\\
$\mathfrak{C}{G}$ massive & 7 & 30 & 19 & 55\\
\end{tabular}
\par\end{centering}
\caption{Counts of master integrals computed with \textsc{Macaulay2}. \label{tab:examples}}
\end{table}

\begin{figure}
\begin{centering}
\includegraphics[scale=0.8]{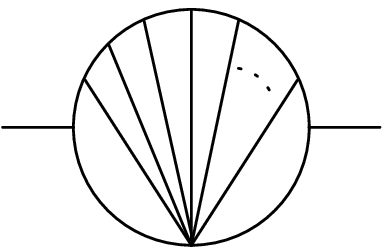}~~~\includegraphics[scale=0.8]{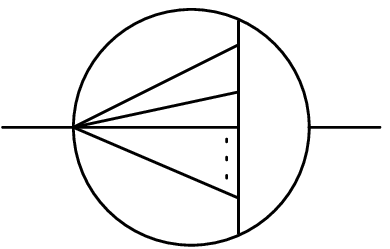}~~~\includegraphics[scale=0.8]{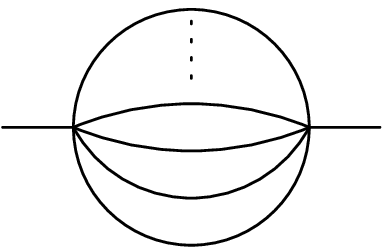}\\
$WS_{L}^{\prime}$~~~~~~~~~~~~~~~~~~~~~~~~~~~~$WS_{L}^{\prime\prime}$~~~~~~~~~~~~~~~~~~~~~~~~~~~~~$S_{L}$
\par\end{centering}
\caption{Three infinite classes of Feynman graphs for which we obtain the number of master integrals in closed form.\label{fig:classes}}
\end{figure}

We have furthermore used identities mentioned above to compute the number of master integrals for the three infinite classes of Feynman graphs indicated in fig. \ref{fig:classes}. The $L-1$-loop graphs $WS^\prime_L$ and $WS^{\prime \prime}$ in fig. \ref{fig:classes} are respectively obtained from cutting a rim or a spoke in the massless wheel with $L$ loops, while the third picture shows the massless sunrise graph $S_L$ with $L$ loops. For the number of master integrals of these graphs we obtain 
\begin{align*}
		\mathfrak{C}\left(WS^\prime_L \right)
		= \mathfrak{C}\left(WS^{\prime \prime}_L \right)
		= \frac{L(L+1)}{2},
		\\
		\mathfrak{C}\left(S_L \right) 
		= 2^{L+1}-1.
\end{align*}
The latter result for the sunrise graphs was previously derived by Kalmykov and Kniehl with a different approach \cite{KalmykovKniehl:CountingMastersSunrise}.

\section{Outlook}\label{sec:Outl}

We have discussed the method of deriving shift relations between Feynman integrals from parametric annihilators of the Lee-Pomeransky polynomial $\mathcal{G}$. As the multivariate Mellin transform bijects these differential operators with the shift relations, the method provides every such relation. In particular it includes the classical IBP relations.
For the number of master integrals, we have introduced an unambiguous definition in terms of the vector space of Feynman integrals associated to a given polynomial $\mathcal{G}$. The number defined in this way is sensitive to all shift relations and does not depend on the method in which they are generated. We have shown that this number is the Euler characteristic of the complement of the hyperspace given by the zero set of $\mathcal{G}$. We have discussed methods to compute this number and provided various examples.

Let us conclude this presentation by pointing out three open questions which arise from this discussion: The first question, most relevant for practical purposes, has to be whether the parametric approach may lead to more efficient reductions to master integrals in the future. We have left this question untouched in our present work. Let us just mention the simple fact that due to the bijection given by the Mellin transform, a desired relation which expresses a given integral in terms of its master integrals obviously corresponds to one particular annihilation operator. Whether the direct search for this operator is a viable alternative to other methods is left for future work. Secondly, having noticed that the IBP relations are contained in the shift relations obtained from parametric annihilators, we may ask whether the reverse is true as well. This would imply, that the classical IBP relations would already provide every shift relation. If this was the case, the ideal of annihilators $\textrm{Ann}_{A^{N}[s]}\left(\mathcal{G}^s\right)$ would be generated by parametric differential operators which correspond to IBP relations. This construction would in most cases be much faster than a direct algorithmic derivation of the ideal. Thirdly, and closely related to the previous question, we ask, whether only linear differential operators are sufficient to generate the ideal $\textrm{Ann}_{A^{N}[s]}\left(\mathcal{G}^s\right)$ for any Lee-Pomeransky polynomial $\mathcal{G}$. If this was the case, the use of syzygies could speed up computations with the annihilation operators. So far, we have studied the second and third question for some graphs of low loop order. In all of these computations, the ideal $\textrm{Ann}_{A^{N}[s]}\left(\mathcal{G}^s\right)$ is generated by linear differential operators and can be obtained from classical IBP relations.

\end{document}